# Fluctuation electromagnetic interaction between rotating spherical nanoparticles. 1. Nonrelativistic theory

A.A. Kyasov and G.V. Dedkov

Nanoscale Physics Group, Kabardino-Balkarian State University, 360004, Nalchik, Russia

For the first time, using nonrelativistic approach we have calculated the attraction force, friction torque and the rate of radiation heat exchange in the system of two sprerical rotating particles located at a distance $R$ between one another.

## 1. Introduction

Manifestation of the fluctuation electromagnetic effects for nanoparticles rotating in photonic gas has recently attracted increasing attention [1,2]. In particular, the particles rotating in empty space (photonic gas filling the vacuum background) acquire the frictional torque, while its thermal radiation spectrum is modified by the frequency of rotation. Similar or even much stronger effects arise when rotating particle is located in the near-field of the surface [3-5], or in the case of the systems of two or more nanoparticles in vacuum. It is worth noting that apart from the aforementioned effects, rotation of the particles also alter the interaction forces (van-der –Waals and Casimir forces). These issues are of paramount importance for the future developments of NEMS, in cluster and cosmic dust physics.

In our preceding papers [3-5] we have considered the fluctuation electromagnetic interaction between a small rotating particle and a solid surface. We have calculated the attraction force, the friction torque and the rate of heat exchange produced by the particle polarization and fluctuating electromagnetic field of the surface. In particular, we have presented a detailed theoretical basis, allowing to calculate the corresponding values in the case of arbitrary orientation of the particle rotation axis relative to the surface. A very similar situation occurs between two rotating particles, one of which can be assumed to be at rest, while another has the rotation velocity $\Omega$. Physically, the particle rotation leads to the frequency shifts in the fluctuating fields involving the frequency $\Omega$. So, the force of attraction (van –der –Waals force) and the rate of heat exchange will be different compared to these quantities in the static case. The friction torque, in its turn, is the net result of rotation. The friction torque is analogous to the dissipative friction force in the case where the particle is in the state of uniform motion near the surface. In this work we want to consider two geometrical configurations, where the rotation axis coincides with or is perpendicular to the radius-vector connecting the particles.

## 2. General results

Figures 1a,b show the geometrical configurations of the systems under consideration. Both spherical particles 1 and 2 (of radii $a_1, a_2$) are assumed to be isotropic, being spaced by the distance $R$. They are characterized by the different temperatures $T_1$, $T_2$ and have the frequency-dependent electric polarizabilities $\alpha_{1,2}(\omega)$. Without loss of generality, we can assume that the second particle is at rest with $\Sigma(X,Y,Z)$ denoting the corresponding reference frame. The first particle rotates with the angular velocity $\Omega$ around the $Z'$ axis (Fig. 1a) that coincides with the $Z$ axis and with the vector $\mathbf{R}$. The case shown in Fig. 1b corresponds to the rotation around the $X'$ axis. The coordinate system $\Sigma'(X',Y',Z')$ rotates with the first particle. Initially, the axes $X', X$ and $Y', Y$ of the systems $\Sigma$ and $\Sigma'$ are parallel one another.

Within the nonrelativistic dipole approximation used in this paper, the problem statement assumes that the following conditions must be fulfilled

$$\max(a_1, a_2) \ll R, \quad \max(a_1, a_2) \ll \min\left\{\frac{2\pi c}{\omega_0}, \frac{2\pi c}{\Omega}, \frac{2\pi \hbar c}{k_B T_1}, \frac{2\pi \hbar c}{k_B T_2}\right\} \tag{1}$$

where $\omega_0$ is the characteristic absorption frequency of the particles, $c, k_B$, and $\hbar$ are the speed of light in vacuum, the Boltzmann constant and Plank's constant.

Following our general method [3--5], we will consider in this case two independent source of electromagnetic fluctuations, namely the spontaneous dipole moments $\mathbf{d}_1^{sp}$ and $\mathbf{d}_2^{sp}$ of particles 1 and 2, taken in the reference system $\Sigma$.

First we calculate the free energy (interaction potential) of the system. The starting expression has the form

$$U = -\frac{1}{2}\left\langle \mathbf{d}_1^{sp}(t)\mathbf{E}_2^{in}(\mathbf{r}_1, t) \right\rangle - \frac{1}{2}\left\langle \mathbf{d}_1^{in}(t)\mathbf{E}_2^{sp}(\mathbf{r}_1, t) \right\rangle = U^{(1)} + U^{(2)} \tag{2}$$

where $\langle ... \rangle$ denotes the total quantum-statistical averaging, the superscripts "sp" and "in" indicate spontaneous and induced quantities taken in the reference frame $\Sigma$, the field $\mathbf{E}_2^{sp}$ is created by spontaneous dipole moment $\mathbf{d}_2^{sp}$ of the second particle. It is worth noting that the fields $\mathbf{E}_2^{in}$ and $\mathbf{E}_2^{sp}$ are taken in the location point of the first particle. In the case shown in Fig.

1a which we first consider, $\mathbf{r}_1 = \mathbf{R} = \mathbf{n} \cdot R = (0,0,R)$. In obtaining the expression for $U^{(1)}$ we use the Fourier transforms (if not

$$\mathbf{d}^{sp}(t) = \int_{-\infty}^{+\infty} \frac{d\omega}{2\pi} \mathbf{d}_1^{sp}(\omega) \exp(-i\omega t) \tag{3}$$

$$\mathbf{E}_2^{in}(\mathbf{r}_1,t) = \int_{-\infty}^{+\infty} \frac{d\omega}{2\pi} \mathbf{E}_2^{in}(\mathbf{r}_1,\omega) \exp(-i\omega t) \tag{4}$$

Correspondingly, the Fourier transform $\mathbf{E}_2^{in}(\mathbf{r}_1,\omega)$ is expressed through the dipole-moment $\mathbf{d}_2^{in}(\omega)$ which is created by the field of the first dipole, namely

$$\mathbf{E}_2^{in}(\mathbf{r}_1,\omega) = \frac{3(\mathbf{d}_2^{in}(\omega) \cdot \mathbf{n}) \cdot \mathbf{n} - \mathbf{d}_2^{in}(\omega)}{R^3} \tag{5}$$

$$\mathbf{d}_2^{in}(\omega) = \alpha_2(\omega) \frac{3(\mathbf{d}_1^{sp}(\omega) \cdot \mathbf{n}) \cdot \mathbf{n} - \mathbf{d}_1^{sp}(\omega)}{R^3} \tag{6}$$

From (5) and (6) we obtain

$$\mathbf{E}_2^{in}(\mathbf{r}_1,\omega) = \frac{\alpha_2(\omega)}{R^6} \left[ 3\mathbf{n} \cdot (\mathbf{d}_1^{sp}(\omega) \cdot \mathbf{n}) + \mathbf{d}_1^{sp}(\omega) \right] \tag{7}$$

Using (2)—(4) and (7) yields (note that we mean the case shown in Fig. 1a)

$$U^{(1)} = -\frac{1}{2(2\pi)^2 R^6} \int_{-\infty}^{+\infty} d\omega' d\omega \exp(-i(\omega+\omega')t) \alpha_2(\omega)$$
$$\left[ 4\langle d_{1z}^{sp}(\omega') d_{1z}^{sp}(\omega) \rangle + \langle d_{1x}^{sp}(\omega') d_{1x}^{sp}(\omega) \rangle + \langle d_{1y}^{sp}(\omega') d_{1y}^{sp}(\omega) \rangle \right]$$
,
(8)

where all components of $\mathbf{d}_1^{sp}(\omega)$ in Eq. (8) are taken in the resting reference frame $\Sigma$. For calculating the correlators in Eq. (8), however, we have to perform the transformation of the corresponding Fourier components from the system $\Sigma$ to the reference frame $\Sigma'$ of the first particle. This is done using the following relations [3]

$$d_x^{sp}(\omega) = \frac{1}{2}\left(d_x^{sp'}(\omega^+) + d_x^{sp'}(\omega^-) + id_y^{sp'}(\omega^+) - id_y^{sp'}(\omega^-)\right)$$

$$d_y^{sp}(\omega) = \frac{1}{2}\left(-id_x^{sp'}(\omega^+) + id_x^{sp'}(\omega^-) + d_y^{sp'}(\omega^+) + d_y^{sp'}(\omega^-)\right) \quad (9)$$

$$d_z^{sp}(\omega) = d_z^{sp'}(\omega)$$

where $\omega^\pm = \omega \pm \Omega$ and the subscript "1" is omitted for brevity. Inserting (9) into (8) we use the fluctuation-dissipation theorem (FDT) for the dipole operator $\mathbf{d}_1^{sp}$ in the rotating coordinate system $\Sigma'$ of the particle

$$\left\langle d_{1i}^{sp'}(\omega')d_{1k}^{sp'}(\omega)\right\rangle = 2\pi\delta_{ik}\delta(\omega+\omega')\hbar\alpha_1''(\omega)\coth\frac{\hbar\omega}{2k_BT_1} \quad (10)$$

As a result we obtain

$$U^{(1)} = -\frac{\hbar}{2\pi R^6}\int_0^\infty d\omega\left[4\alpha_2'(\omega)\alpha_1''(\omega)\coth\frac{\hbar\omega}{2k_BT_1} + \alpha_2'(\omega)\alpha_1''(\omega^+)\coth\frac{\hbar\omega^+}{2k_BT_1} + \right.$$

$$\left. + \alpha_2'(\omega)\alpha_1''(\omega^-)\coth\frac{\hbar\omega^-}{2k_BT_1}\right] \quad (11)$$

Now we pass to the calculation of $U^{(2)}$ which is given by

$$U^{(2)} = -\frac{1}{2}\left\langle \mathbf{d}_1^{in}(t)\mathbf{E}_2^{sp}(\mathbf{r}_1,t)\right\rangle = -\frac{1}{2}\int_{-\infty}^{+\infty}\frac{d\omega'}{2\pi}\frac{d\omega}{2\pi}\exp(-i(\omega+\omega')t)\left\langle \mathbf{d}_1^{in}(\omega')\mathbf{E}_2^{sp}(\mathbf{r}_1,\omega)\right\rangle \quad (12)$$

$$\mathbf{E}_2^{sp}(\mathbf{r}_1,\omega) = \frac{3(\mathbf{d}_2^{sp}(\omega)\cdot\mathbf{n})\cdot\mathbf{n} - \mathbf{d}_2^{sp}(\omega)}{R^3} \quad (13)$$

When substituting (13) into (12) we must express the components of $\mathbf{d}_1^{in}(\omega)$ through the field $\mathbf{E}_2^{sp}(\mathbf{r}_1,\omega)$. This is done in two stages. First, we transform $\mathbf{E}_2^{sp}(\mathbf{r}_1,t)$ to the rotating system $\Sigma'$ of the first particle and use the relation $\mathbf{d}_1^{in'}(\omega) = \alpha_2(\omega)\mathbf{E}_2^{sp'}(\mathbf{r}_1,\omega)$ (see [3] in more detail). Second, we transform the dipole moment $\mathbf{d}_1^{in'}(\omega)$ from $\Sigma'$ to $\Sigma$. The resulting expressions for the components of $\mathbf{d}_1^{in}(\omega)$ have the form

$$d^{in}_x(\omega) = \frac{1}{2}\left[\alpha(\omega^+)\left(E^{sp}_x(\omega) + iE^{sp}_y(\omega)\right) + \alpha(\omega^-)\left(E^{sp}_x(\omega) - iE^{sp}_y(\omega)\right)\right]$$

$$d^{in}_y(\omega) = \frac{1}{2}\left[-i\alpha(\omega^+)\left(E^{sp}_x(\omega) + iE^{sp}_y(\omega)\right) + i\alpha(\omega^-)\left(E^{sp}_x(\omega) - iE^{sp}_y(\omega)\right)\right], \quad (14)$$

$$d^{in}_z(\omega) = \alpha(\omega)E^{sp}_z(\omega)$$

where we have omitted for brevity the subscripts "1" for components of the dipole moment, and "2" for the polarizability and components of the electric field. Making use (13)--(14), the correlators of the dipole moments $d^{sp}_{1i}(\omega)$ ($i = x, y, z$) in the reference system $\Sigma$ are calculated through the FDT of the form

$$\left\langle d^{sp}_{2i}(\omega')d^{sp}_{2k}(\omega)\right\rangle = 2\pi\delta_{ik}\delta(\omega+\omega')\hbar\alpha''_2(\omega)\coth\frac{\hbar\omega}{2k_BT_2} \quad (15)$$

Finally, using (12)—(15) we obtain

$$U^{(2)} = -\frac{\hbar}{2\pi R^6}\int_0^\infty d\omega\left[4\alpha''_2(\omega)\alpha'_1(\omega)\coth\frac{\hbar\omega}{2k_BT_2} + \alpha''_2(\omega)\alpha'_1(\omega^+)\coth\frac{\hbar\omega}{2k_BT_2} + \right.$$
$$\left. + \alpha''_2(\omega)\alpha'_1(\omega^-)\coth\frac{\hbar\omega}{2k_BT_2}\right] \quad (16)$$

Summing (11) and (16) yields

$$U = -\frac{\hbar}{2\pi R^6}\int_0^\infty d\omega\left[4\left(\alpha''_1(\omega)\alpha'_2(\omega)\coth\frac{\hbar\omega}{2k_BT_1} + \alpha'_1(\omega)\alpha''_2(\omega)\coth\frac{\hbar\omega}{2k_BT_2}\right) + \right.$$
$$+ \alpha''_1(\omega^+)\alpha'_2(\omega)\coth\frac{\hbar\omega^+}{2k_BT_1} + \alpha'_1(\omega^+)\alpha''_2(\omega)\coth\frac{\hbar\omega}{2k_BT_2} + \quad (17)$$
$$\left. + \alpha''_1(\omega^-)\alpha'_2(\omega)\coth\frac{\hbar\omega^-}{2k_BT_1} + \alpha'_1(\omega^-)\alpha''_2(\omega)\coth\frac{\hbar\omega}{2k_BT_2}\right]$$

The starting equations in the calculations of the friction torque $M_z$ and the rate of heating (cooling) of the first particle are given by

$$M_z = \left\langle\left(\mathbf{d}^{sp}_1(t)\times\mathbf{E}^{in}_2(\mathbf{r}_1,t)\right)_z\right\rangle + \left\langle\left(\mathbf{d}^{in}_1(t)\times\mathbf{E}^{sp}_2(\mathbf{r}_1,t)\right)_z\right\rangle \quad (18)$$

$$\dot{Q} = \left\langle \mathbf{d}_1^{sp}(t)\mathbf{E}_2^{in}(\mathbf{r}_1,t) \right\rangle + \left\langle \mathbf{d}_1^{in}(t)\mathbf{E}_2^{sp}(\mathbf{r}_1,t) \right\rangle \tag{19}$$

The calculations in (18), (19) are absolutely the same as the calculation of $U$. The resulting formulas are

$$M_z = -\frac{\hbar}{\pi R^6} \int_0^\infty d\omega\, \alpha_2''(\omega) \left[ \begin{array}{l} \alpha_1''(\omega^-)\left(\coth\dfrac{\hbar\omega^-}{2k_B T_1} - \coth\dfrac{\hbar\omega}{2k_B T_2}\right) - \\ -\alpha_1''(\omega^+)\left(\coth\dfrac{\hbar\omega^+}{2k_B T_1} - \coth\dfrac{\hbar\omega}{2k_B T_2}\right) \end{array} \right] \tag{20}$$

$$\dot{Q} = \frac{\hbar}{\pi R^6} \int_0^\infty d\omega\, \omega\, \alpha_2''(\omega) \left[ 4\alpha_1''(\omega)\left(\coth\dfrac{\hbar\omega}{2k_B T_2} - \coth\dfrac{\hbar\omega}{2k_B T_1}\right) + \right.$$
$$\left. +\alpha_1''(\omega^+)\left(\coth\dfrac{\hbar\omega}{2k_B T_2} - \coth\dfrac{\hbar\omega^+}{2k_B T_1}\right) + \alpha_1''(\omega^-)\left(\coth\dfrac{\hbar\omega}{2k_B T_2} - \coth\dfrac{\hbar\omega^-}{2k_B T_1}\right) \right] \tag{21}$$

In the case shown in Fig. 1b the calculations are performed according to the same scheme. The difference is related to Eqs. (9) and (14) that should be modified in this case using the cyclic permutation of the indices $x \to y, y \to z, z \to x$ [5]. The resulting expressions for $U, M_x$ and $\dot{Q}$ have the form

$$U = -\frac{\hbar}{4\pi R^6} \int_0^\infty d\omega \left[ 2\left(\alpha_1''(\omega)\alpha_2'(\omega)\coth\dfrac{\hbar\omega}{2k_B T_1} + \alpha_1'(\omega)\alpha_2''(\omega)\coth\dfrac{\hbar\omega}{2k_B T_2}\right) + \right.$$
$$+ 5\left(\alpha_1''(\omega^+)\alpha_2'(\omega)\coth\dfrac{\hbar\omega^+}{2k_B T_1} + \alpha_1'(\omega^+)\alpha_2''(\omega)\coth\dfrac{\hbar\omega}{2k_B T_2}\right) +$$
$$\left. + 5\left(\alpha_1''(\omega^-)\alpha_2'(\omega)\coth\dfrac{\hbar\omega^-}{2k_B T_1} + \alpha_1'(\omega^-)\alpha_2''(\omega)\coth\dfrac{\hbar\omega}{2k_B T_2}\right) \right] \tag{22}$$

$$M_x = -\frac{5\hbar}{2\pi R^6} \int_0^\infty d\omega\, \alpha_2''(\omega) \left[ \begin{array}{l} \alpha_1''(\omega^-)\left(\coth\dfrac{\hbar\omega^-}{2k_B T_1} - \coth\dfrac{\hbar\omega}{2k_B T_2}\right) - \\ -\alpha_1''(\omega^+)\left(\coth\dfrac{\hbar\omega^+}{2k_B T_1} - \coth\dfrac{\hbar\omega}{2k_B T_2}\right) \end{array} \right] \tag{23}$$

$$\dot{Q} = \frac{\hbar}{2\pi R^6} \int_0^\infty d\omega\, \omega \alpha_2''(\omega) \left[ 2\alpha_1''(\omega) \left( \coth\frac{\hbar\omega}{2k_B T_2} - \coth\frac{\hbar\omega}{2k_B T_1} \right) + \right.$$
$$\left. + 5\alpha_1''(\omega^+) \left( \coth\frac{\hbar\omega}{2k_B T_2} - \coth\frac{\hbar\omega^+}{2k_B T_1} \right) + 5\alpha_1''(\omega^-) \left( \coth\frac{\hbar\omega}{2k_B T_2} - \coth\frac{\hbar\omega^-}{2k_B T_1} \right) \right]$$
(24)

One can easily see that at $\Omega = 0$, $T_1 = T_2 = 0$ Eqs. (17) and (22) agree with the non-retarded expression for the van –der –Waals energy [6], while Eqs. (21) and (24) at $\Omega = 0$ agree with the non-retarded expression for the rate of heat exchange [7].

## 3. Estimation of stopping times of nanoparticles

Similar to [3—5], let us compare the stopping times of the rotating nanoparticles in free vacuum, in the near field of the surface, and in the case of interaction between two particles with the same radius $a$ and material properties. As an example, we have chosen the material properties of SiC (for particles and for the surface)

$$\varepsilon(\omega) = \varepsilon_\infty + \frac{\omega_T^2 (\varepsilon_0 - \varepsilon_\infty)}{\omega_T^2 - \omega^2 - i\gamma\omega} \tag{25}$$

Correspondingly, the particle polarizability is given by

$$\alpha(\omega) = a^3 \frac{(\varepsilon(\omega) - 1)}{(\varepsilon(\omega) + 2)} \tag{26}$$

Parameters of $SiC$ correspond to [8]: $\varepsilon_0 = 9.8$, $\varepsilon_\infty = 6.7$, $\omega_T = 0.098\, eV$, $\gamma = 0.00585\, eV$. We also assume the case of thermal equilibrium with the temperature $T$. Equation (23) then takes the form

$$M_x = -\frac{5\hbar\Omega}{2\pi R^6} \int_{-\infty}^{+\infty} d\omega (\alpha''(\omega))^2 \left(-\frac{d}{d\omega}\right) \coth\frac{\hbar\omega}{2k_B T} \tag{27}$$

In the case of particle rotation near the surface (with the rotation axis being parallel to the surface) the friction torque is given by [5] ($z_0$ is the particle-surface distance)

$$M_x = -\frac{3\hbar\Omega}{16\pi z_0^3} \int_{-\infty}^{+\infty} d\omega\, \mathrm{Im}\left(\frac{\varepsilon(\omega) - 1}{\varepsilon(\omega) + 2}\right) \alpha''(\omega) \left(-\frac{d}{d\omega}\right) \coth\frac{\hbar\omega}{2k_B T} \tag{28}$$

Moreover, when the particle rotates in free vacuum (photonic gas with the temperature $T$), the friction torque has the form [1]

$$M_x^{(vac)} = -\frac{2\hbar\Omega}{3\pi c^3}\int_{-\infty}^{+\infty}d\omega\,\omega^3\,\text{Im}\left(\frac{\varepsilon(\omega)-1}{\varepsilon(\omega)+2}\right)\alpha''(\omega)\left(-\frac{\partial}{\partial\omega}\right)\coth\frac{\hbar\omega}{2k_B T} \qquad (29)$$

Using (27)—(29) and Newton's second law, we obtain $\Omega(t) = \Omega(0)\exp(-t/\tau)$ where $\tau$ is the characteristic decay time of rotation. Bearing in mind the inertia moment of spherical particle $8\pi\rho a^5/15$, the stopping times corresponding to (28)—(30) will be given by

$$\begin{aligned}
\tau_{P-P} &= \frac{16\pi^2}{75}\rho\frac{R^6}{a\hbar J_{P-P}(\omega_W)} \\
\tau_{P-S} &= \frac{128\pi^2}{45}\frac{\rho a^5}{\hbar}\frac{z_0^3}{J_{P-S}(\omega_W)}) \\
\tau_{vac} &= \frac{4\pi^2}{5}\frac{\rho a^5}{\hbar}\frac{c^3}{J_{vac}(\omega_W)}
\end{aligned} \qquad (30)$$

where $J_{P-P}, J_{P-S}, J_{vac}$ denote the frequency integrals in (27)—(29) and $\omega_W = k_B T/\hbar$. Figure 1 compares the calculated stopping times at $T = 300K, a = 1nm, \rho = 3.22 g/cm^3$. One can see that at small separations $\tau_{P-P} \approx \tau_S \ll \tau_{vac}$, but $\tau_{P-P}$ increases much faster than $\tau_S$ with increasing the distance.

## 4. Conclusions

Using the fluctuation electromagnetic theory, we have obtained closed nonrelativistic expressions for the friction torque, van –der –Waals energy and heating rate of a small spherical particle rotating close to the other particle. We have considered two geometrical configurations where the rotation axis coincides with or is perpendicular to the radius vector between the particles. Material properties of the particles are characterized by the frequency-dependent polarizabilities, while the temperatures are assumed to be arbitrary. Apart from the distance dependence, the obtained expressions for the quantities under consideration are very similar to the previously obtained in the case of a particle rotating in the near field of the surface. At small separations between the particles the friction torque (between SiC particles) turns out to be by about eight orders of magnitude larger than in the case of a particle rotating in free vacuum. A very intriguing question concerns the impact of rotation on the value of the van –der –Waals force, but it needs special analysis.

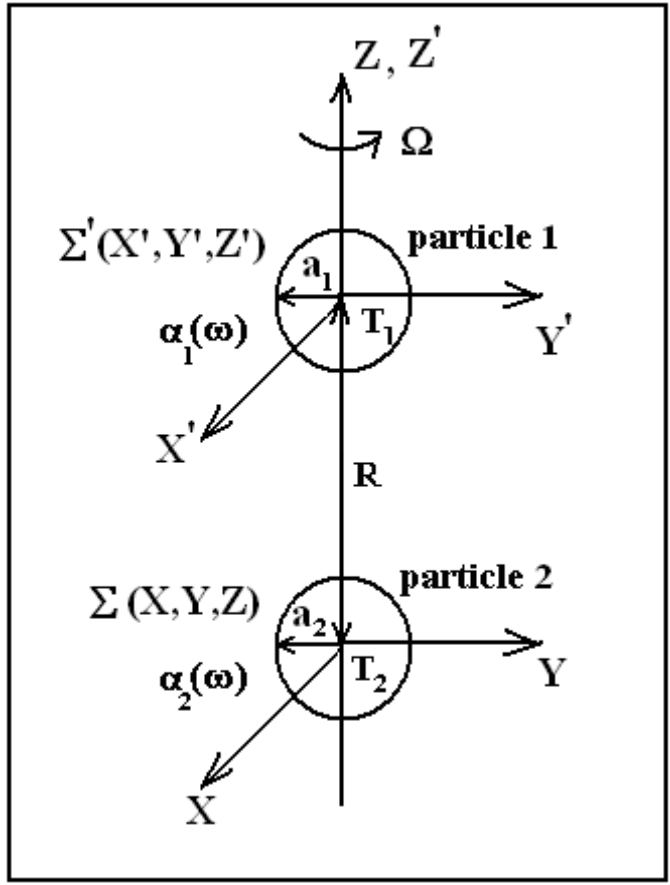

Fig. 1a Configuration 1 and coordinate systems used

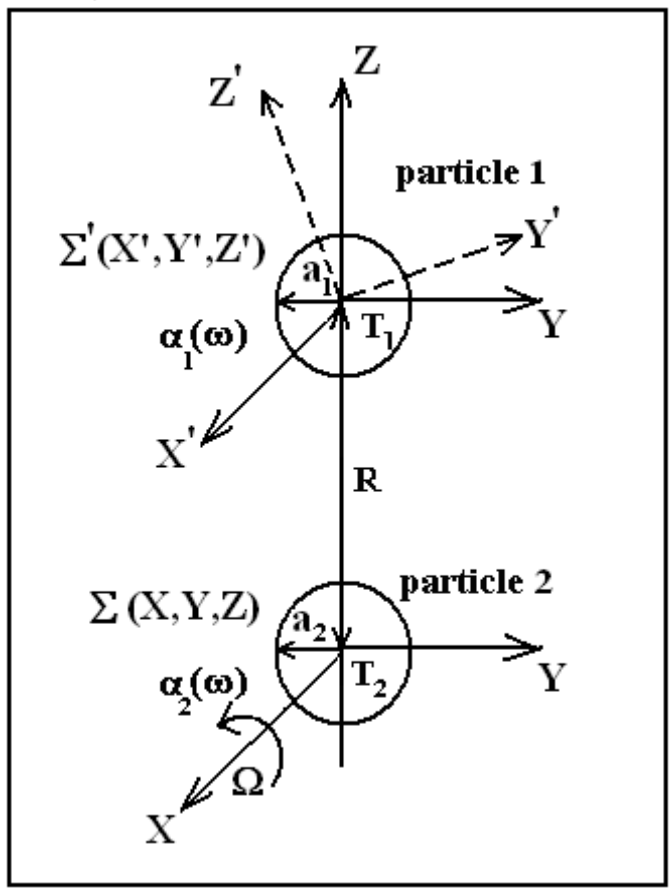

Fig. 1b Configuration 2 and coordinate systems used

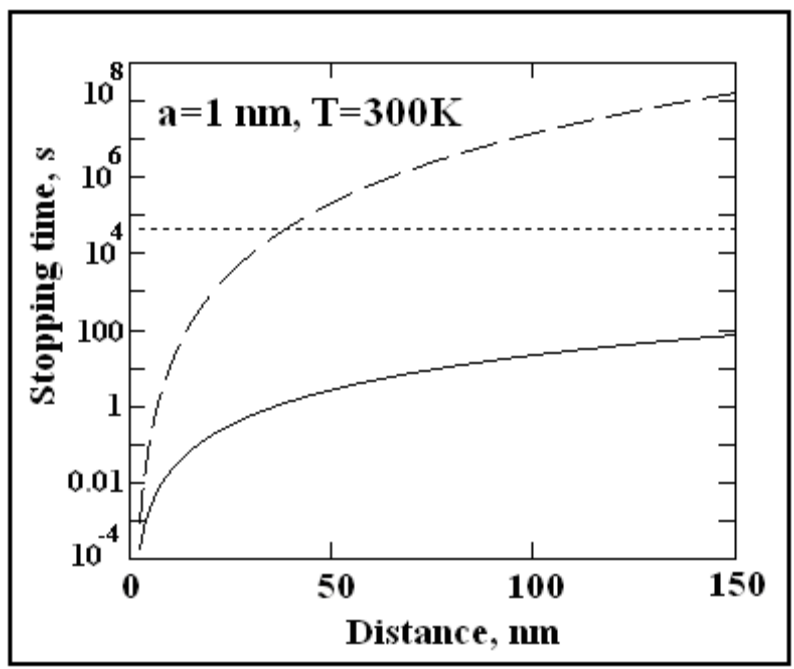

Fig. 2 Stopping times of SiC particles, corresponding to the rotation near the surface of SiC (solid line), in free vacuum (dotted line), and in the system of two SiC particles (dashed line).